\documentclass[aps,prl,reprint,superscriptaddress]{revtex4-2}

\usepackage{graphicx}
\usepackage{xcolor}
\usepackage{dcolumn}
\usepackage{bm, gensymb, siunitx, xfrac, physics}
\usepackage{hyperref}
\usepackage{booktabs, multirow}

\setcitestyle{numbers, round}
\hypersetup{
    colorlinks=true,
    urlcolor= blue,
    citecolor=blue,
    linkcolor= blue,
    bookmarksopen=false,
    }
    
\newif\ifptitle
\newif\ifpnumber
\newcounter{para}
\newcommand\ptitle[1]{\par\refstepcounter{para}
{\ifpnumber{\noindent\textcolor{darkgray}{\textbf{\thepara}}\indent}\fi}
{\ifptitle{\textbf{[{#1}]}}\fi}}
\pnumbertrue  

\definecolor{sblue}{RGB}{33, 118, 199}
\definecolor{mblue}{RGB}{0, 118, 186}
\definecolor{hblue}{RGB}{36, 34, 111}

\definecolor{mgreen}{RGB}{29, 177, 0}

\definecolor{sred}{RGB}{209, 28, 36}

\definecolor{orang}{RGB}{255, 147, 0}

\definecolor{gry}{RGB}{100, 100, 100}

\begin{document}

\title{Visualizing the atomic-scale origin of metallic behavior in Kondo insulators}

\author{Harris Pirie}
\affiliation{Department of Physics, Harvard University, Cambridge, MA 02138, USA}
\affiliation{Clarendon Laboratory, University of Oxford, Oxford, OX1 3PU, UK}
\author{Eric Mascot}
\affiliation{Department of Physics, University of Illinois at Chicago, Chicago, IL 60607, USA}
\author{Christian E. Matt}
\author{Yu Liu}
\author{Pengcheng Chen}
\author{M. H. Hamidian}
\affiliation{Department of Physics, Harvard University, Cambridge, MA 02138, USA}
\author{Shanta Saha}
\author{Xiangfeng Wang}
\author{Johnpierre Paglione}
\affiliation{Maryland Quantum Materials Center, Department of Physics, University of Maryland, College Park, MD 20742, USA}
\author{Graeme Luke}
\affiliation{Department of Physics and Astronomy, McMaster University, Hamilton, ON L8S 4M1, Canada}
\author{David Goldhaber-Gordon}
\affiliation{Department of Physics, Stanford University, Stanford, CA 94305, USA}
\affiliation{Stanford Institute for Materials and Energy Sciences, SLAC National Accelerator Laboratory, Menlo Park, CA 94025, USA}
\author{Cyrus F. Hirjibehedin}
\altaffiliation[Present address: ]{Lincoln Laboratory, Massachusetts Institute of Technology, Lexington, MA 02421-6426, USA.}
\affiliation{London Centre for Nanotechnology, University College London (UCL), London WC1H 0AH, UK}
\affiliation{Department of Physics and Astronomy, UCL, London WC1E 6BT, UK}
\affiliation{Department of Chemistry, UCL, London WC1H 0AJ, UK}
\author{J. C. S\'{e}amus Davis}
\affiliation{Clarendon Laboratory, University of Oxford, Oxford, OX1 3PU, UK}
\affiliation{Department of Physics, University College Cork, Cork T12 R5C, Ireland}
\affiliation{LASSP, Department of Physics, Cornell University, Ithaca, NY 14850, USA}
\affiliation{Max-Planck Institute for Chemical Physics of Solids, D-01187 Dresden Germany}
\author{Dirk K. Morr}
\affiliation{Department of Physics, University of Illinois at Chicago, Chicago, IL 60607, USA}
\author{Jennifer E. Hoffman}
\email{jhoffman@physics.harvard.edu}
\affiliation{Department of Physics, Harvard University, Cambridge, MA 02138, USA}

\date{\today}

\begin{abstract}
A Kondo lattice is often electrically insulating at low temperatures. However, several recent experiments have detected signatures of bulk metallicity within this Kondo insulating phase. Here we visualize the real-space charge landscape within a Kondo lattice with atomic resolution using a scanning tunneling microscope. We discover nanometer-scale puddles of metallic conduction electrons centered around uranium-site substitutions in the heavy-fermion compound URu$_2$Si$_2$, and around samarium-site defects in the topological Kondo insulator SmB$_6$. These defects disturb the Kondo screening cloud, leaving behind a fingerprint of the metallic parent state. Our results suggest that the mysterious 3D quantum oscillations measured in SmB$_6$ could arise from these Kondo-lattice defects, although we cannot rule out other explanations. Our imaging technique could enable the development of atomic-scale charge sensors using heavy-fermion probes.  
\end{abstract}

\nocite{Doniach1977, Steglich1979, Andres1975, Fisk1986, Laurita2016, Flachbart2006, Fuhrman2018, Thomas2019, Hartstein2018, Tan2015, Millichamp2021, Eo2019, Knolle2015, Baskaran2015, Erten2017, Shen2018, Skinner2019, Fuhrman2020, Abele2020, Dagotto2005, Sollie1991, Figgins2011, Figgins2019, Hamidian2011, Urbano2007, Nefedova1999, Gabani2002, Orendac2017, Souza2020, Valentine2016, Eo2021, Koslowski1995, Wagner2015, Hapala2016, Nonnenmacher1991, GrossScience2009, Mohn2012, Zerweck2005, Sadewasser2009, AlbrechtPRL2015, AlbrechtPRB2015, Kotta2022, Madhavan1998, Maltseva2009, Figgins2010, Schmidt2010, Giannakis2019, Kohsaka2007, delaTorre1992, Santander-Syro2009, Matt2020, Sun2018, Pirie2020, Jiang2013, Jiao2018, Akintola2017, Liu2009, Okada2011, Jack2020, Phelan2016, Bork2011, Aishwarya2022, Zenodo}

\maketitle

 \begin{figure*}
	\includegraphics[width=5.6in]{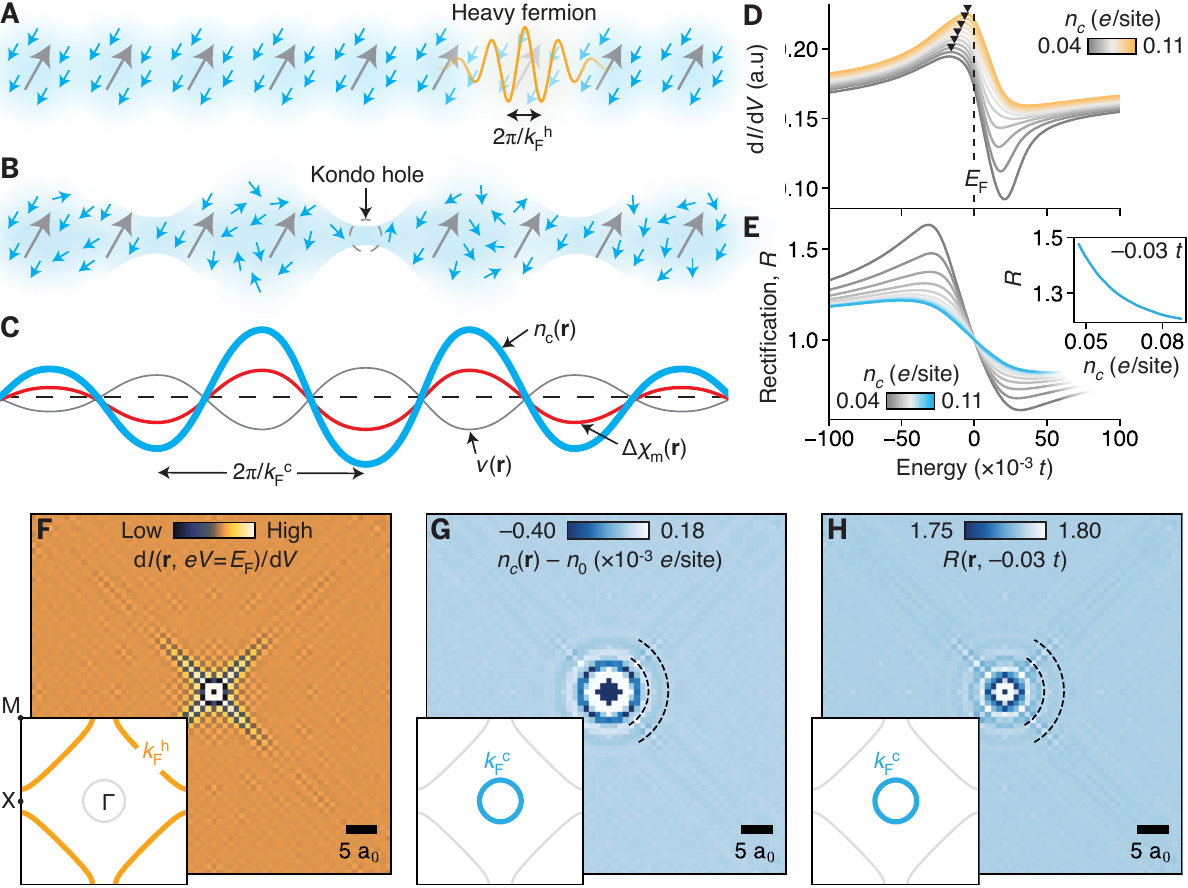}
	\caption{\label{fig:1}{\bf Expected disruption of the screening cloud around Kondo holes.}
	({\bf A}) In a uniform Kondo lattice, magnetic moments at each site (gray arrows) are coherently screened by itinerant conduction electrons (blue cloud) to form a spinless ground state of heavy fermions (orange line), characterized by the wavevector $k_\mathrm{F}^\mathrm{h}$. 
	({\bf B}) If one moment is removed to create a Kondo hole, the conduction electrons previously screening it can redistribute themselves. 
	({\bf C}) The redistributed screening cloud causes oscillations of the local conduction electron density $n_\mathrm{c}(\mathbf{r})$, interaction strength $\nu(\mathbf{r})$, and magnetic susceptibility $\chi_\mathrm{m}(\mathbf{r})$ at the conduction-band wavevector $k_\mathrm{F}^\mathrm{c}$, as shown schematically. 
	({\bf D}) In a uniform Kondo lattice, the Kondo resonance creates a peak-dip feature in the calculated d$I$/d$V$, caused by the quantum interference between tunneling into the conduction band and the $f$-electron states with respective amplitudes $t_c$ and $t_f$. The energy position of the peak (black triangles) shifts linearly according to the local conduction-electron density $n_\mathrm{c}$. 
	({\bf E}) The calculated rectification $R(V) = |I(+V)/I(-V)|$ acquires a strong peak because d$I$/d$V$ is asymmetric around the Fermi level $E_F$ (which occurs at $V=0$). The $R(V)$ peak amplitude depends on the d$I$/d$V$ peak energy. These changes are almost linear over the small range of local doping expected around a Kondo hole (inset).
	({\bf F}) The calculated oscillations in d$I(\mathbf{r}, V)$/d$V$ at the Fermi level around a Kondo hole match the hybridized Fermi surface ($k_\mathrm{F}^\mathrm{h}$, orange line in inset). 
	({\bf G}) In contrast, the calculated $n_c(\mathbf{r})$ varies according to the circular wavevector of the unhybridized Fermi surface ($k_\mathrm{F}^\mathrm{c}$, blue line in inset) as it mainly reflects the disturbance to the screening cloud. 
	({\bf H}) Calculated $R(\mathbf{r}, V)$ is dominated by unhybridized electrons for biases within the hybridization gap. 
    The calculations in (D)-(H) are based on a Kondo-Heisenberg model with nearest-neighbor hopping strength $t$, Kondo coupling $J=2t$, antiferromagnetic exchange $I=0.002t$, and tunneling amplitudes $t_f/t_c=-0.025$. In (D) and (E), the hybridization strength is fixed at $\nu=0.1 t$ and the antiferromagnetic correlation strength is fixed at $\chi=0.0003t$. The Fermi wavelength is $\lambda_F^c = 8a_0$ in (F)-(H), where $a_0$ is the lattice spacing.
	}
\end{figure*}

\begin{figure*}
	\includegraphics[width=5.6in]{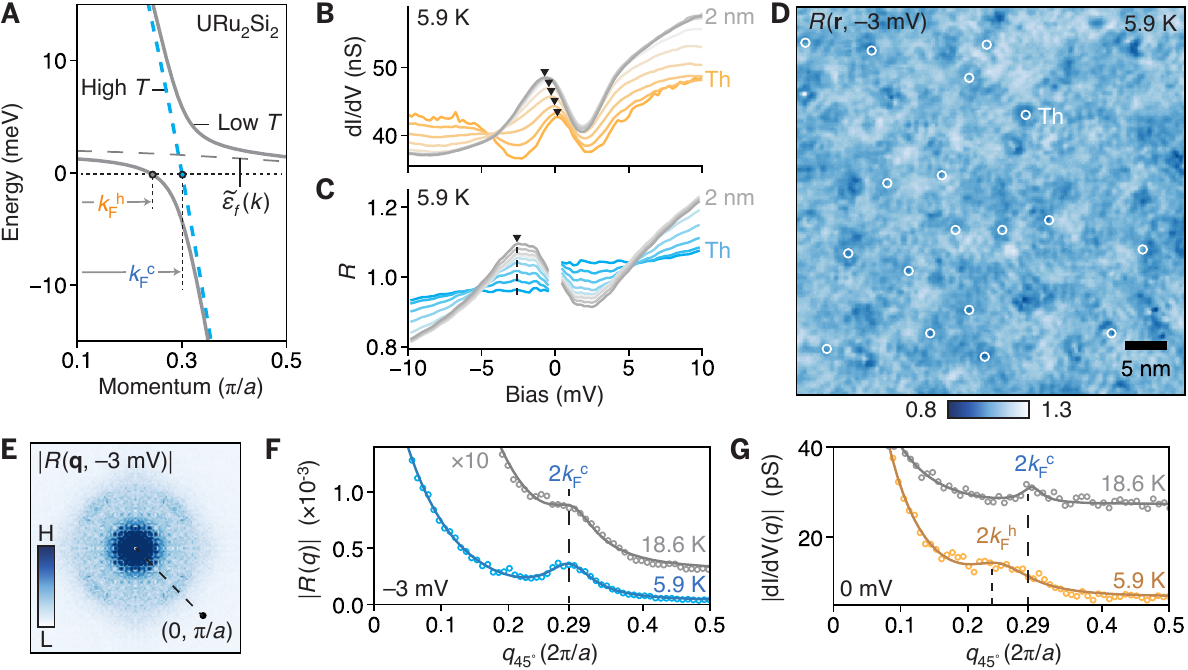}
	\caption{\label{fig:2}{\bf Thorium dopants induce Kondo-hole behavior in URu$_2$Si$_2$.}
	({\bf A}) Schematic band structure of URu$_2$Si$_2$ showing the onset of heavy fermion bands (gray solid lines) at temperatures below $T_\mathrm{o}=17.5$ K, as itinerant conduction electrons (blue line) hybridize with a renormalized 5$f$ level (gray dashed line), reducing the Fermi wavevector from $k_\mathrm{F}^\mathrm{c}$ to $k_\mathrm{F}^\mathrm{h}$. 
	({\bf B}) Experimental measurement of an asymmetric Fano lineshape in the tunneling conductance at temperatures below $T_o$ on the U termination (gray curve). This feature shifts towards the Fermi level near a thorium dopant (black triangles), consistent with an expected change in local charge density. 
	({\bf C}) For a fixed bias, the $R(V)$ peak amplitude (black triangle) is highly sensitive to the d$I$/d$V$ peak position. The spectra in (B) and (C) are averaged over the 18 well-isolated thorium dopants marked in (D). 
	({\bf D}) The measured $R(\mathbf{r},V)$ exhibits clear oscillations that manifest as a ring in ({\bf E}) the 4-fold-symmetrized Fourier transform. 
	({\bf F}) These oscillations match the high-temperature Fermi wavevector of $2k_\mathrm{F}^\mathrm{c} \approx 0.3\ (2\pi/a)$, both above and below $T_\mathrm{o}$. 
	({\bf G}) In contrast, a conventional d$I$/d$V$ measurement couples to the temperature-dependent Fermi surface, which changes dramatically from 18.6 K to 5.9 K. For clarity, the 18.6 K data have been scaled in (F) and offset in (G). 
	}
\end{figure*}

\ptitle{Remnant metallicity in Kondo lattices}
When the electrons in a material interact strongly with one another, they often produce unexpected behavior. Above a characteristic temperature $T_K$, a lattice of local $f$ moments within a conducting Fermi sea behaves like an ordinary magnetic metal, with a Curie-Weiss susceptibility. But below $T_K$, the competition between antiferromagnetic ordering of the local moments and their screening by conduction electrons leads to a rich phase diagram, exhibiting quantum criticality \cite{Doniach1977}, unconventional superconductivity \cite{Steglich1979}, and heavy fermions \cite{Andres1975, Fisk1986}---quasiparticles with $f$-electron character (Fig.~\ref{fig:1}A). A Kondo insulator forms if the spectral gap opened by hybridization between the conduction band and the renormalized $f$ band spans the Fermi level. Mysteriously, some Kondo insulators seem to remember their metallic parent state long after this gap is fully developed. For example, the topological Kondo insulator SmB$_6$ displays a sizable bulk optical conductivity at terahertz frequencies \cite{Laurita2016} and a finite electronic specific heat at low temperatures \cite{Flachbart2006, Fuhrman2018, Thomas2019, Hartstein2018}. A complete 3D Fermi surface matching its high-temperature metallic state was reconstructed from quantum oscillation \cite{Tan2015, Hartstein2018} and Compton scattering \cite{Millichamp2021} measurements performed in the insulating regime. Notably, these metallic properties persist even as the bulk resistivity of SmB$_6$ increases by 10 orders of magnitude \cite{Eo2019}. This discrepancy led to several theoretical proposals: some argue that the metallic behavior is intrinsic, either a consequence of the small hybridization gap in Kondo insulators \cite{Knolle2015}, or arising from exotic charge-neutral quasiparticles \cite{Baskaran2015, Erten2017}; others suggest an extrinsic origin \cite{Shen2018, Skinner2019, Fuhrman2020, Abele2020}, which implies the presence of microscopic metallic pockets.

\ptitle{Kondo holes nucleate metallic puddles}
Charge inhomogeneity is commonplace at nanometer length scales, especially in materials with strong electron interactions that promote competing orders \cite{Dagotto2005}. In a Kondo lattice, defects that substitute or remove the $f$-contributing moment, called Kondo holes, have a widespread impact on the nearby electronic structure \cite{Sollie1991, Figgins2011, Figgins2019}. First, these defects locally untangle the hybridized wavefunction, leaving puddles of unhybridized conduction electrons behind (Fig.~\ref{fig:1}B). In theory, these charge puddles should have the same itinerant character as the metallic parent state \cite{Figgins2011}, but they have not been imaged directly. Additionally, the excess conduction electrons released from hybridization adjust the strength of their interactions with the remaining $f$ moments \cite{Figgins2011, Hamidian2011}, leading to enhanced local magnetism \cite{Urbano2007} (Fig.~\ref{fig:1}C). For example, Sm$_{1-x}$La$_x$B$_6$ samples with nonmagnetic La dopants are known to display increased specific heat and magnetic susceptibility compared to undoped samples \cite{Nefedova1999, Gabani2002, Orendac2017}. More recently, the existence of local metallic puddles around Gd dopants in Sm$_{1-x}$Gd$_x$B$_6$ was inferred from electron spin-resonance measurements \cite{Souza2020}. Meanwhile, an increased concentration of Sm vacancies in Sm$_{1-x}$B$_6$ was shown to globally inhibit the development of the hybridization gap \cite{Valentine2016}, eventually leading to bulk conduction \cite{Eo2019, Eo2021}. All of these findings suggest that Sm-site defects manifest as Kondo holes in SmB$_6$, yet their key signature---the accompanying charge oscillations relating to the parent metallic Fermi surface \cite{Figgins2011}---remain undetected by any microscopic probe.

\ptitle{These charge puddles are challenging to measure.}
Directly imaging the metallic puddles around Kondo holes is difficult, because the inherent screening strongly renormalizes the bare charge distribution. However, there are a few promising approaches \cite{Koslowski1995, Wagner2015, Hapala2016}. The most common is to decorate the tip of a Kelvin probe force microscope with a single atom or molecule \cite{Nonnenmacher1991, GrossScience2009}. This technique was used to image the charge variations within an adsorbed molecule \cite{Mohn2012}. However, it becomes inaccurate for small tip-sample separations because of the influence of short-range forces \cite{Zerweck2005, Sadewasser2009}, complicating further improvements to its spatial resolution \cite{AlbrechtPRL2015}. Meanwhile, a scanning tunneling microscope (STM) routinely achieves the sub-nanometer spatial resolution, cryogenic temperatures, and sub-meV energy resolution required to access atomic charge distributions, but existing methods to extract the electrostatic potential from the STM vacuum decay length contain significant artifacts \cite{AlbrechtPRB2015}. Consequently, simultaneously achieving the high charge precision and high spatial resolution required to measure the charge environment around a Kondo hole is not possible using existing techniques.

\begin{figure*}
	\includegraphics[width=5.6in]{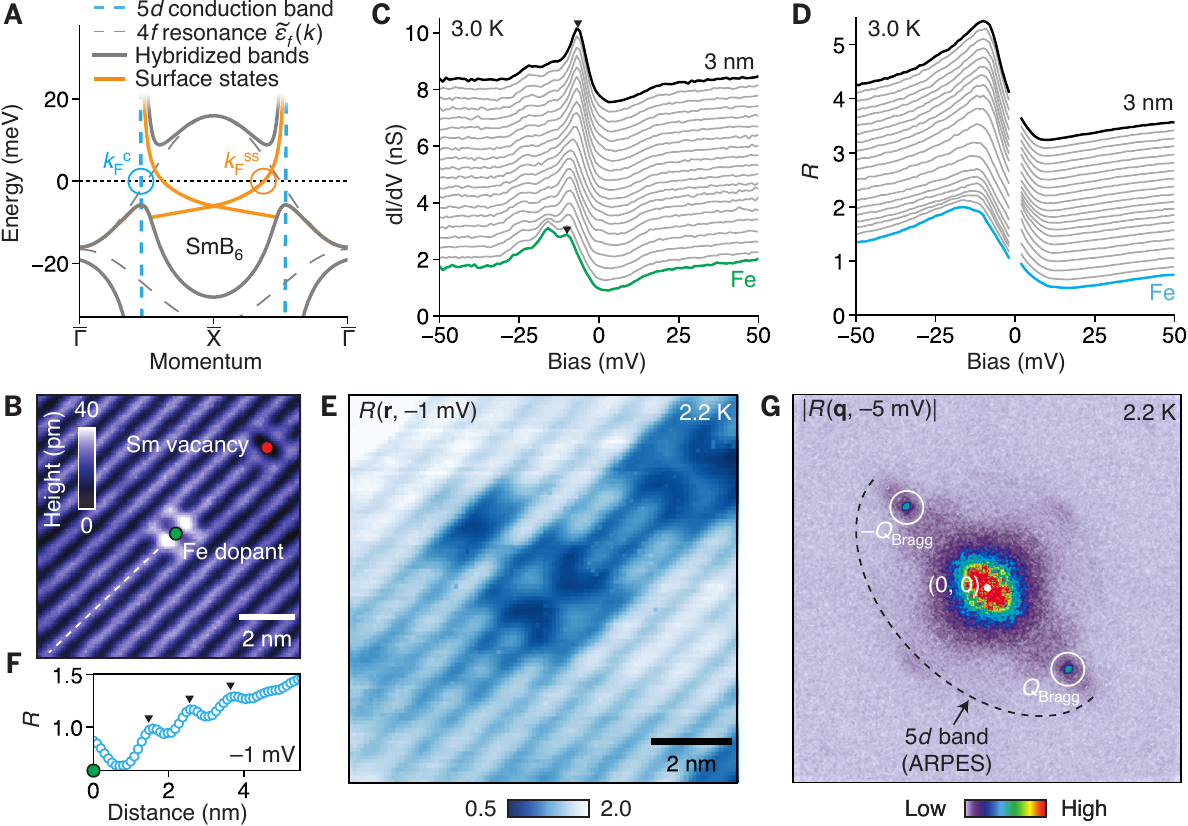}
	\caption{\label{fig:3}{\bf Kondo holes nucleate metallic puddles in SmB$_6$.}
	({\bf A}) Schematic band structure of SmB$_6$ showing the hybridization between conduction electrons (blue dashed line) and localized 4$f$ moments (gray dashed line), which leads to an inverted band structure (gray solid line) hosting emergent heavy Dirac surface states with a reduced Fermi wavevector (orange). 
	({\bf B}) STM topography of the $(2 \times 1)$-reconstructed Sm surface of lightly Fe-doped SmB$_6$. Both the Fe dopant and Sm vacancy in this image are expected to act as Kondo holes because they each displace a 4$f$ moment. 
	({\bf C-D}) Near the Fe dopant, the measured d$I$/d$V$ peak changes energy position (black triangles), leading to large variations in $R(\mathbf{r}, V)$ peak amplitude. The spectra in (C) and (D) have been offset for clarity. 
	({\bf E}) $R(\mathbf{r}, V)$ in the same area as shown in (B) contains clear oscillations around the two impurities. 
	({\bf F}) Linecut of $R(\mathbf{r}, V)$ along the white dashed line in (B). 
	({\bf G}) $R(\mathbf{r}, V)$ oscillations appear as a sharp ring in the 2-fold-symmetrized Fourier transform (taken from a larger $65 \times 80$-nm$^2$ area for enhanced $\mathbf{q}$ resolution), which matches the unhybridized 5$d$ Fermi surface inferred from ARPES experiments \cite{Jiang2013} (dashed line). The surface reconstruction creates a sharp peak in $R(\mathbf{r}, V)$ at $Q_\mathrm{Bragg} = (0, \pi/a)$.
	}
\end{figure*}

\ptitle{Here we experimentally image the metallic puddles around Kondo holes.}
Here we develop a dedicated STM modality to image the charge environment within a Kondo lattice with sub-\aa{}ngstr\"{o}m resolution. At temperatures below $T_K$, we image charge oscillations matching the parent Fermi surface centered around spinless thorium atoms in the Kondo metal URu$_2$Si$_2$, and around three separate Sm-site defects in the Kondo insulator SmB$_6$. Importantly, the charge puddles we image in SmB$_6$ exhibit the same metallic wavevector seen in recent quantum oscillation experiments \cite{Tan2015, Hartstein2018}, suggesting Kondo-lattice defects are the source of those oscillations.

\subsection{Measuring local charge density in a Kondo lattice}

\ptitle{The charge around a Kondo hole modulates $\tilde{\varepsilon}_f(\mathbf{r})$.}
To visualize the conduction-electron density $n_c(\mathbf{r})$ in a Kondo lattice (and hence local charge $-e n_c(\mathbf{r})$, where $-e$ is the electron charge), we first show theoretically that $n_c(\mathbf{r})$ determines the energy position of the Kondo resonance $\tilde{\varepsilon}_f(\mathbf{r})$, which forms near the Fermi level as the magnetic $f$ moments are screened by conduction electrons. Then, we establish an experimental metric capable of detecting the sub-meV variations in $\tilde{\varepsilon}_f(\mathbf{r})$ around a Kondo hole. Our technique takes advantage of how the many-body Kondo resonance responds to local doping. In the Abrikosov fermion representation for local moments, $\tilde{\varepsilon}_f$ is the Lagrange multiplier that enforces uniform $f$-electron density, typically $n_f=1$ at each site. As additional charge carriers $\Delta n_c$ enter a uniform Kondo lattice, the hybridized Fermi surface reshapes to accommodate them, leading to a corresponding change in $\tilde{\varepsilon}_f$ in order to maintain $n_f = 1$  (see black triangles in Fig.~\ref{fig:1}D, and Fig.~S2E). The magnitude and direction of the shift in $\tilde{\varepsilon}_f$ depend on the details of the band structure. But the relationship between $n_c$ and $\tilde{\varepsilon}_f$ is linear over a wide range of band parameters and charge doping (see Fig.~S2), implying that the charge density at position $\mathbf{r}$, can usually be inferred by measuring $\tilde{\varepsilon}_f(\mathbf{r})$. In fact, the linear dependence of $\tilde{\varepsilon}_f(\mathbf{r})$ on $n_c(\mathbf{r})$ was recently verified experimentally by micron-scale angle-resolved photoemission (ARPES) measurements in Eu-doped SmB$_6$ \cite{Kotta2022}.

\ptitle{Rectification amplifies shifts in $\tilde{\varepsilon}_f$}
In STM measurements, the Kondo resonance normally appears as a peak-dip feature in the tunneling conductance d$I$/d$V$ \cite{Madhavan1998} (where $I$ is the sample-to-tip tunneling current at applied sample bias $V$), because of the presence of multiple tunneling channels \cite{Maltseva2009, Figgins2010} (see calculation in Fig.~\ref{fig:1}D). In simple cases, $\tilde{\varepsilon}_f$ can be estimated by fitting d$I$/d$V$ to a Fano-like model \cite{Schmidt2010, Giannakis2019}. However, the exact value of $\tilde{\varepsilon}_f$ depends on the model used, so this approach is not immediately suitable for detecting the small, sub-meV energy shifts in $\tilde{\varepsilon}_f (\mathbf{r})$ expected around a Kondo hole. Instead, we track the ratio of forward-to-backward tunneling current, i.e.~the local rectification $R(\mathbf{r}, V) = |I(\mathbf{r}, +V)/I(\mathbf{r}, -V)|$. This ratio is insensitive to STM setup artifacts, and it was previously used to track charge inhomogeneity from the spectral weight transfer at high biases in hole-doped cuprates \cite{Kohsaka2007}. Here, we focus on low biases, typically $V \lesssim 10$ mV, where the small shifts in $\tilde{\varepsilon}_f (\mathbf{r})$ generate large variations in $R(\mathbf{r}, V)$ owing to the energy asymmetry of d$I$/d$V$ about the Fermi level at $V=0$ (see calculations in Fig.~\ref{fig:1}, D and E, and Fig.~S2). To demonstrate this effect locally, we self-consistently calculated d$I(\mathbf{r}, V)$/d$V$, $n_c(\mathbf{r})$, and $R(\mathbf{r}, V)$ around a Kondo hole in a metallic Kondo lattice, as shown in Fig.~\ref{fig:1}, F to H. The calculated d$I(\mathbf{r}, V)$/d$V$ at $V=0$ tracks the local Fermi-level density of states, so it reveals the hybridized Fermi surface of heavy fermions with a wavevector $2 k_\mathrm{F}^\mathrm{h}$. In contrast, both $n_c(\mathbf{r})$ and $R(\mathbf{r}, V)$ are dominated by static oscillations at the unhybridized wavevector $2 k_\mathrm{F}^\mathrm{c}$, associated with the Friedel-like redistribution of the Kondo screening cloud. The correlation between $n_c(\mathbf{r})$ and $R(\mathbf{r}, V)$ establishes $R(\mathbf{r}, V)$ as a qualitative probe of local charge, except at very short distances from a Kondo hole ($|\mathbf{r}| \sim a$) likely because $n_f = 1$ is not enforced at that site.

\subsection{Kondo holes in URu$_2$Si$_2$}

\ptitle{In URu$_2$Si$_2$, $R(\mathbf{r},V)$ reveals widespread $2k_\mathrm{F}^\mathrm{c}$ oscillations emanating from Th dopants.}
To test our technique, we first studied the Kondo metal URu$_2$Si$_2$ with 1\% thorium dopants, which are known to induce Kondo-hole behavior \cite{delaTorre1992, Hamidian2011}. Previous STM measurements mapped a metal-like Fermi surface in URu$_2$Si$_2$ for temperatures above $T_\mathrm{o} = 17.5$ K, consisting of a single conduction band with wavevector $k_\mathrm{F}^\mathrm{c} \approx0.3\ \pi/a$, where $a$ is the lattice constant (\cite{Schmidt2010}, see Fig.~2A). The onset of coherent heavy fermion bands below $T_\mathrm{o}$ \cite{Santander-Syro2009} is accompanied by the appearance of a peak-dip feature in d$I$/d$V$, i.e.~the Kondo-Fano resonance (see Fig.~\ref{fig:2}B). Close to a thorium dopant, this feature shifts upwards in energy, towards the Fermi level. This energy shift---and even the barely perceptible shifts 2 nm away from the dopant---are easily detected in the amplitude of $R(\mathbf{r}, V)$ (see Fig.~\ref{fig:2}C). For biases within the hybridization gap $|V| < \Delta/e \approx 5$ mV (where $\Delta$ is the gap magnitude), $R(\mathbf{r}, V)$ displays widespread spatial oscillations emanating from thorium dopants, as shown in Fig.~\ref{fig:2}, D to F. Their wavevector of $0.29 \pm 0.01\ (2\pi/a)$ agrees perfectly with the hybridization oscillations previously measured around Kondo holes in this compound \cite{Hamidian2011}. It matches the URu$_2$Si$_2$ parent metallic Fermi surface detected above $T_\mathrm{o}$ from our measured quasiparticle interference patterns in d$I(\mathbf{r}, V)$/d$V$ at $V=0$, but it is distinct from the heavy bands that we measured below $T_\mathrm{o}$ (see Fig.~\ref{fig:2}G). As a final check, we independently extracted $\tilde{\varepsilon}_f(\mathbf{r})$ by fitting d$I(\mathbf{r}, V)$/d$V$ curves to a Fano model (see Fig.~S3). The excellent agreement between $\tilde{\varepsilon}_f(\mathbf{r})$ and $R(\mathbf{r}, V)$ corroborates the existence of charge oscillations at $2k_\mathrm{F}^\mathrm{c}$ in URu$_2$Si$_2$, indicating that some electrons retain their itinerant character around Kondo holes, even below $T_\mathrm{o}$.

\begin{figure}
	\includegraphics[width=2.72in]{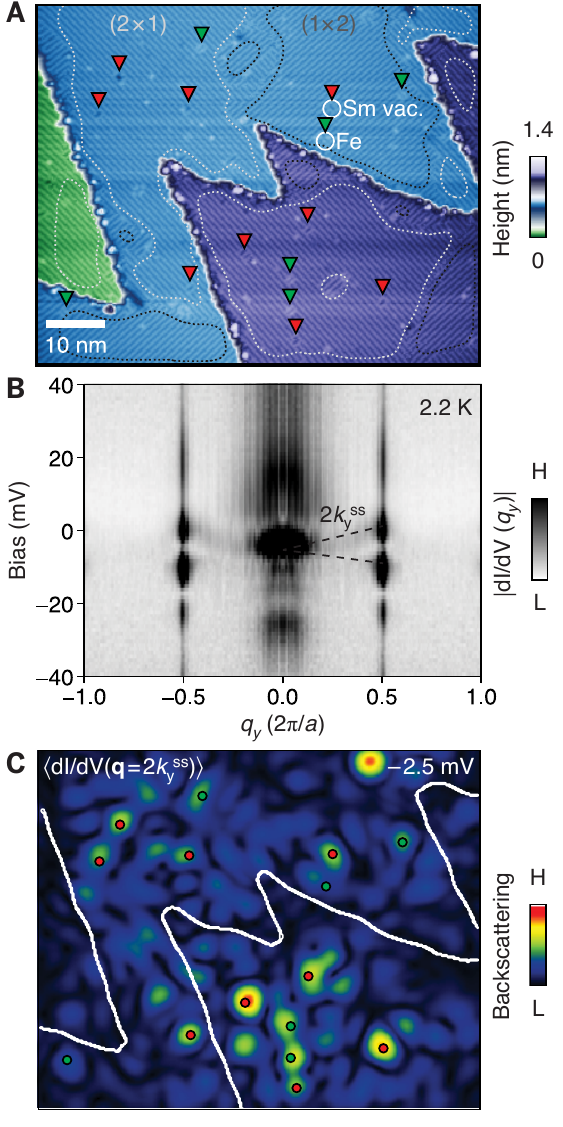}
	\caption{\label{fig:4}{\bf Kondo holes backscatter heavy Dirac fermions.}
    ({\bf A}) Topography of an SmB$_6$ region that contains 15 well-isolated Kondo holes (position indicated by red and green triangles) on several $(2 \times 1)$- or $(1 \times 2)$-reconstructed domains (dotted lines).
    ({\bf B})  For energies within the Kondo-insulating gap, the Fourier-transformed d$I$/d$V$ along $q_\mathrm{y}$ (perpendicular to Sm rows) contains a linearly dispersing signal (black dashed line) corresponding to quasiparticle interference from backscattered heavy Dirac fermions. The Fourier transform from the $(1 \times 2)$ domains was rotated by 90$^\circ$ before being averaged with that from the $(2 \times 1)$ domains.
    ({\bf C}) The intensity of backscattering from topological states, calculated from Fourier-filtering d$I$/d$V$ at the $y$ component of the backscattering wavevector $q_y=2k_y^{\mathrm{ss}}$, is strongly peaked around each Kondo hole. This map is computed only for ordered patches of the sample, as marked in (A), and excludes step edges. 
	}
\end{figure}

\subsection{Metallic puddles in SmB$_6$}

\ptitle{The $R(\mathbf{r},V)$ oscillations in SmB$_6$ reveal light-band electrons around Kondo holes}
In our Kondo insulating SmB$_6$ samples, any atomic defect that replaces a Sm atom to alter the $4f$ moment could generate metallic puddles like those seen in URu$_2$Si$_2$. We searched for these puddles in flux-grown samples lightly doped with Fe, which contain two clear Sm-site defects: Sm vacancies and Fe substitutions (see Fig.~\ref{fig:3}B). We focused on the $(2 \times 1)$ Sm termination, as its charge environment most closely represents that of the bulk \cite{Matt2020}. As in URu$_2$Si$_2$, we noticed that the d$I$/d$V$ peak attributed to the Kondo resonance changes its energy position near candidate Kondo holes (Fig.~\ref{fig:3}C), strongly impacting the $R(\mathbf{r}, V)$ peak amplitude (Fig.~\ref{fig:3}D). Similar shifts in d$I$/d$V$ peak position were previously linked to the buildup of charge around boron clusters on the Sm $(1\times 1)$ termination \cite{Sun2018}. For biases within the hybridization gap $|V| < \Delta/e \approx 10$ mV, $R(\mathbf{r}, V)$ reveals prominent oscillations around Sm-site defects (see Fig.~\ref{fig:3}, E and F). These oscillations create a sharp ellipse in the Fourier transform of $R(\mathbf{r}, V)$, as shown in Fig.~\ref{fig:3}G. The wavevectors of the $R(\mathbf{q}, V)$ ellipse are larger than those of the surface state detected by quasiparticle interference imaging \cite{Pirie2020} and they do not disperse for biases within the hybridization gap, indicating a different origin (see Fig.~S4). On the other hand, the size and shape of the ellipse matches the unhybridized $5d$ band found by extrapolating ARPES data \cite{Jiang2013} to the Fermi level (i.e.~it matches the SmB$_6$ metallic parent state), after accounting for band folding on the $(2 \times 1)$ surface (see Fig.~\ref{fig:3}G and Fig.~S5). Our observation of this $5d$ wavevector within the Kondo insulating gap is direct evidence of atomic-scale metallicity around Kondo holes. This metallicity is supported by the large residual d$I$/d$V$ at $V=0$ mV that we measured around Kondo holes (green curve in Fig.~\ref{fig:3}C), indicating a sizable Fermi-level density of states even when the metallic surface states are suppressed \cite{Jiao2018}. We confirmed this discovery by checking for $R(\mathbf{r}, V)$ oscillations around a third type of Kondo hole, Gd dopants, as detailed in Fig.~S6.

\subsection{Magnetic fluctuations at Kondo holes in SmB$_6$}

\ptitle{Kondo holes harbor magnetic fluctuations.}
Our $R(\mathbf{r}, V)$ maps show the real-space structure of the metallic puddles around Kondo holes in SmB$_6$. For these puddles to contribute to the measured de Haas-van Alphen oscillations in magnetization, they must have a finite magnetic susceptibility. Several Sm-site defects are already suspected to be locally magnetic based on their impact on bulk susceptibility \cite{Nefedova1999, Gabani2002, Akintola2017, Fuhrman2018} and their influence on the topologically emergent surface states \cite{Pirie2020, Jiao2018}. In general, topological surface states can provide a test of local magnetism because they are protected against backscattering from non-magnetic defects, but not from magnetic defects that locally break time-reversal symmetry \cite{Liu2009}. This additional magnetic backscattering was previously imaged around Fe dopants in two Bi-based topological insulators \cite{Okada2011, Jack2020}. Here we visualize the intensity of magnetic fluctuations at Sm-site defects in SmB$_6$ by identifying spatial regions where its surface states backscatter. For biases within the hybridization gap, we measured large-area d$I(\mathbf{r}, V)$/d$V$ maps that contain clear quasiparticle interference patterns at the backscattering wavevector $\mathbf{q}\equiv\mathbf{k}_\mathrm{f}-\mathbf{k}_\mathrm{i}=2\mathbf{k}^{\mathrm{ss}}$  (Fig.~\ref{fig:4}B), consistent with our previous report \cite{Pirie2020}. We determined the spatial origin of this signal by Fourier-filtering d$I(\mathbf{r}, V)$/d$V$ at the wavevector $2\mathbf{k}^{\mathrm{ss}}$ to create an image of the local backscattering strength (Fig.~\ref{fig:4}C). Most of the peaks in this image align with the positions of Sm vacancies or Fe dopants, indicating that these Kondo holes harbor the necessary magnetic fluctuations to backscatter topological states.

\subsection{Discussion and Outlook}

\ptitle{Charge puddles explain the metallicity.}
The charge puddles around Kondo holes present an alternative yet compelling origin for many of the strange observations of metallic behavior in SmB$_6$. First, the detection of de Haas-van Alphen (magnetic) oscillations without accompanying Shubnikov-de Haas (resistivity) oscillations \cite{Tan2015, Hartstein2018, Thomas2019} is expected for electrically isolated metallic puddles, provided they do not meet the percolation threshold (which could be unreachable \cite{Skinner2019}). Second, the large Fermi surface size and light effective mass extracted by bulk probes \cite{Tan2015, Hartstein2018, Millichamp2021} is in excellent agreement with our observation of itinerant $5d$ electrons. Third, the magnetic length of the high-frequency (large-$k_F$) quantum oscillations that onset above 35 T \cite{Tan2015, Hartstein2018} is comparable to the $R(\mathbf{r})$ decay length of $\gamma =2.6$ nm, such that a Landau orbit could fit inside a metallic puddle.  Additionally, many of the metallic properties were detected in floating-zone-grown samples \cite{Tan2015, Flachbart2006, Hartstein2018, Millichamp2021, Laurita2016}, which are known to have higher concentrations of Sm vacancies than samples grown with an aluminum flux \cite{Phelan2016}. Floating-zone samples also contain a higher concentration of dislocations \cite{Eo2021}, which may similarly disrupt the Kondo screening cloud and thus further enhance the quantum oscillation amplitude beyond that expected from Sm vacancies alone. In contrast, the quantum oscillations completely disappear in flux-grown samples once embedded aluminum is removed \cite{Thomas2019}.

\ptitle{Conclusion}
Atomic-scale charge inhomogeneity has a profound impact on many interacting quantum materials, but it has typically not been possible to measure. In Kondo-lattice systems, $R(\mathbf{r},V)$ provides a peek at the ground-state charge landscape, which is strongly perturbed by Kondo holes. These Kondo holes nucleate nanometer-scale metallic puddles that could explain many of the strange phenomena detected by bulk probes. More broadly, the sensitivity to local charge within a Kondo lattice may enable atomic-scale charge imaging using STM tips decorated with a Kondo impurity \cite{Bork2011} or fabricated from heavy-fermion materials \cite{Aishwarya2022}.

\ptitle{Acknowledgements} {\bf Acknowledgements} We thank An-Ping Li, Brian Skinner, Christian Wagner, Felix L\"{u}pke, Stefan Ulrich, Yun Suk Eo, and Zachary Fisk, for helpful conversations. We thank Anjan Soumyanarayanan, Michael Yee, and Yang He for their help measuring Gd-doped SmB$_6$. 
{\bf Funding:} This project was supported by the Gordon and Betty Moore Foundation's EPiQS Initiative through grants GBMF4536, GBMF9071, and GBMF9457. The experiments at Harvard were supported by the US National Science Foundation grant DMR-1410480. The data interpretation received support from AFOSR grant FA9550-21-1-0429. The work of E.M.~and D.K.M.~was supported by the US Department of Energy, Office of Science, Basic Energy Sciences, under Award DE-FG02-05ER46225. C.E.M.~is supported by the Swiss National Science Foundation under fellowship P400P2\_183890. Work at the University of Maryland was supported by AFOSR FA9550-22-1-0023. Research at McMaster University was supported by the Natural Sciences and Engineering Research Council. J.C.S.D.~acknowledges support from the Science Foundation of Ireland under Award SFI 17/RP/5445, and from the European Research Council under Award DLV-788932. This project received funding from the European Union's Horizon 2020 research and innovation programme under the Marie Sk\l{}odowska-Curie grant agreement 893097. 
{\bf Author contributions:} H.P., C.E.M., Y.L., P.C., and M.H.H.~carried out the STM experiments. S.S., X.W., J.P., and G.L.~synthesized the samples. E.M.~and D.K.M.~developed the theoretical model. D.G.G., C.F.H., and J.C.S.D.~contributed to the understanding of the results. H.P.~and J.E.H.~analyzed the data and wrote the manuscript with contributions from E.M., C.F.H., J.C.S.D., and D.K.M.
{\bf Competing interests:} The authors have no competing interests. 
{\bf Data and materials availability:} All data and analysis presented in this paper are deposited in Zenodo \cite{Zenodo}.

\bibliography{refs}

\end{document}


\title{{\Large Supplementary Information} \\ \vspace{0.20cm} Visualizing the atomic-scale origin of metallic behavior in Kondo insulators}

\author{Harris Pirie}
\author{Eric Mascot}
\author{Christian E. Matt}
\author{Yu Liu}
\author{Pengcheng Chen}
\author{M. H. Hamidian}
\author{Shanta Saha}
\author{Xiangfeng Wang}
\author{Johnpierre Paglione}
\author{Graeme Luke}
\author{David Goldhaber-Gordon}
\author{Cyrus Hirjibehedin}
\author{J. C. S\'{e}amus Davis}
\author{Dirk K. Morr}
\author{Jennifer E. Hoffman}

\maketitle

\section{Materials and Methods}

\nocite{Doniach1977, Steglich1979, Andres1975, Fisk1986, Laurita2016, Flachbart2006, Fuhrman2018, Thomas2019, Hartstein2018, Tan2015, Millichamp2021, Eo2019, Knolle2015, Baskaran2015, Erten2017, Shen2018, Skinner2019, Fuhrman2020, Abele2020, Dagotto2005, Sollie1991, Figgins2011, Figgins2019, Hamidian2011, Urbano2007, Nefedova1999, Gabani2002, Orendac2017, Souza2020, Valentine2016, Eo2021, Koslowski1995, Wagner2015, Hapala2016, Nonnenmacher1991, GrossScience2009, Mohn2012, Zerweck2005, Sadewasser2009, AlbrechtPRL2015, AlbrechtPRB2015, Kotta2022, Madhavan1998, Maltseva2009, Figgins2010, Schmidt2010, Giannakis2019, Kohsaka2007, delaTorre1992, Santander-Syro2009, Matt2020, Sun2018, Pirie2020, Jiang2013, Jiao2018, Akintola2017, Liu2009, Okada2011, Jack2020, Phelan2016, Bork2011, Aishwarya2022, Zenodo}

\subsection{STM measurements}
\begin{figure}[h!]
	\includegraphics[width=5in]{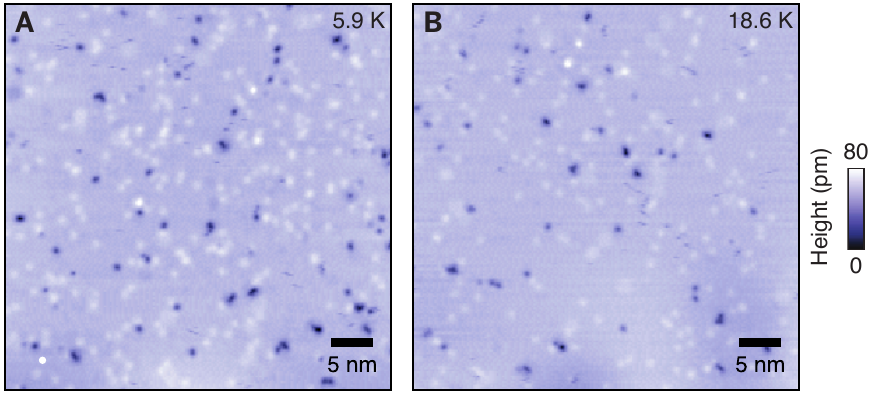}
		\caption{{\bf Topographies for URu$_2$Si$_2$ measurements.} \label{fig:URS-topo}
    	Topographies of the $50 \times 50$ nm$^2$ regions of the U-terminated surface of URu$_2$Si$_2$ where data was collected at 
        ({\bf A}) 5.9 K and
        ({\bf B}) 18.6 K. Dark spots correspond to surface Th atoms. White spots correspond to subsurface impurities.
        }
\end{figure}

\ptitle{} We studied single crystals of 1\% Th-doped URu$_2$Si$_2$ grown using the Czochralski method, and 0.1\% Gd-doped and 0.5\% Fe-doped SmB$_6$ grown using the Al-flux method \cite{Kim2013}. The measured doping concentration in the STM topographies was typically lower than the nominal doping, as shown in Table \ref{tab:doping}. Each sample was cleaved in cryogenic ultrahigh vacuum and immediately inserted into a variable-temperature scanning tunneling microscope (STM). Due to the cryogenic vacuum conditions, atomically flat terraces were stable for many months without noticeable degradation. The corresponding topographies for the temperature-dependent measurements on URu$_2$Si$_2$ are shown in Fig.~\ref{fig:URS-topo}. The d$I(\mathbf{r},V)$/d$V$ measurements were acquired with a standard alternating current lock-in amplifier technique. STM tips were cut from PtIr wire and cleaned by {\it in situ} field emission on Au foil.

\subsection{Theory}
\ptitle{Eric's description of calculations}The calculations in Fig.~1 are based on the Kondo-Heisenberg Hamiltonian,
\begin{align}\label{eq:hamiltonian}
	\mathcal{H} = \sum_{\vec{k}, \sigma} \varepsilon_{\vec{k}} c_{\vec{k}, \sigma}^\dag c_{\vec{k}, \sigma}
	+ J \sum_{\vec{r}} \vec{S}_{\vec{r}} \cdot \vec{s}_{\vec{r}}
	+ I \sum_{\braket{\vec{r}, \vec{r}'}} \vec{S}_{\vec{r}} \cdot \vec{S}_{\vec{r}'},
\end{align}
where the conduction electrons have the dispersion \(\varepsilon_{\vec{k}} = -2t(\cos k_x + \cos k_y) - \mu\) with nearest-neighbor hopping, \(t\), and chemical potential, \(\mu\);
\(c_{\vec{k}, \sigma}^\dag\) (\(c_{\vec{k}, \sigma}\)) creates (annihilates) a conduction electron with momentum \(\vec{k}\) and spin \(\sigma\);
\(J\) is the Kondo coupling between the local \(f\) moment, \(\vec{S}_{\vec{r}}\), and the conduction electron's spin, \(\vec{s}_{\vec{r}} = \frac{1}{2} \sum_{\alpha, \beta} c_{\vec{r}, \alpha}^\dag \vec{\sigma}_{\alpha\beta} c_{\vec{r}, \beta}\); and
\(I\) is the antiferromagnetic Heisenberg exchange coupling.
We proceed on the interpretation that a Kondo hole at site $\mathbf{r}_0$ is described by removing all terms in Eq.~\ref{eq:hamiltonian} that include $\mathbf{S}_{\mathbf{r}_0}$ and adding an onsite electrostatic potential $U \sum_\sigma c_{\vec{r}_0 \sigma}^\dagger c_{\vec{r}_0 \sigma}$. We set $U = t$ for the calculations in Fig.~1, F to H.
We use a spin-1/2 Abrikosov fermion representation \cite{Abrikosov1965} for the \(f\) moments, \(\vec{S}_{\vec{r}} = \frac{1}{2} \sum_{\alpha, \beta} f_{\vec{r}, \alpha}^\dag \vec{\sigma}_{\alpha\beta} f_{\vec{r}, \beta}\) subject to the constraint \(n_f(\vec{r}) = \sum_\sigma f_{\vec{r}, \sigma}^\dag f_{\vec{r}, \sigma} = 1\).
We enforce the constraint using a Lagrange multiplier for each site, \(\tilde{\varepsilon}_f(\vec{r})\).
We introduce the mean-fields,
\begin{align}
	\nu(\vec{r}) =& -\frac{J}{2} \sum_\sigma \braket{f_{\vec{r}, \sigma}^\dag c_{\vec{r} \sigma}}, &
	\chi(\vec{r}, \vec{r}') =& \frac{I}{2} \sum_\sigma \braket{f_{\vec{r}', \sigma}^\dag f_{\vec{r}, \sigma}},
\end{align}
where $\nu(\vec{r})$ is the hybridization and $\chi(\vec{r}, \vec{r}')$ is the $f$-electron dispersion, to obtain the mean-field Hamiltonian,
\begin{align}\label{eq:Hfull}
	\mathcal{H} =
	\sum_{\vec{k}, \sigma} \varepsilon_{\vec{k}} c_{\vec{k}, \sigma}^\dag &c_{\vec{k}, \sigma}
	- \sum_{\vec{r}, \sigma} \tilde{\varepsilon}_f(\vec{r}) f_{\vec{r}, \sigma}^\dag f_{\vec{r}, \sigma} \nonumber 
	 -\sum_{\braket{\vec{r}, \vec{r}'}, \sigma} \left(
		\chi(\vec{r}, \vec{r}') f_{\vec{r}, \sigma}^\dag f_{\vec{r}', \sigma}
		+ \text{H.c.}
	\right) \nonumber \\
	& + \sum_{\vec{r}, \sigma} \left(
		\nu(\vec{r}) c_{\vec{r}, \sigma}^\dag f_{\vec{r}, \sigma}
		+ \text{H.c.}
	\right),
\end{align}
where H.c. is the Hermitian conjugate. 
The Hamiltonian in Eq.~\ref{eq:Hfull} was diagonalized in real space on a $64 \times 64$ square lattice with periodic boundary conditions, allowing a self-consistent calculation of the mean-field parameters $\tilde{\varepsilon}_f(\mathbf{r})$, $\nu(\mathbf{r})$, and $\chi(\mathbf{r}, \mathbf{r}^\prime)$ following the procedure described in Ref.~\cite{Johnson1988}.
To compute the tunneling conductance \(\mathrm{d}I/\mathrm{d}V\), we introduce the tunneling term
\begin{align}
	\mathcal{H}_T = \sum_{\sigma} \left(
		t_c d_{\vec{R}, \sigma}^\dag c_{\vec{R}, \sigma}
		+ t_f d_{\vec{R}, \sigma}^\dag f_{\vec{R}, \sigma}
		+ \mathrm{H.c.}
	\right),
\end{align}
where \(d_{\vec{R}, \sigma}^\dag\) creates an electron in the STM tip.
The tunneling conductance in the zero-temperature, wide-band, weak-tunneling limit is then given by
\begin{align}
	\frac{\mathrm{d}I(\vec{r}, V)}{\mathrm{d}V} \propto
	t_c^2 N_c(\vec{r}, eV) + t_f^2 N_f(\vec{r}, eV) + 2 t_c t_f N_{cf}(\vec{r}, eV),
\end{align}
where \(N_c, N_f\) are the local density of states of the conduction electrons and \(f\) electrons, respectively, and \(N_{cf} = -\mathrm{Im} G_{cf} / \pi\) \cite{Figgins2010}. The current is given by     
\begin{align}
        I(\mathbf{r}, V) = \int_0^V dV’ \frac{dI(\mathbf{r}, V’)}{dV’} .
\end{align}
Finally, the rectification $R(\mathbf{r}, V)$ is then computed from 
\begin{align}
    R(\mathbf{r}, V) = \qty|\frac{I(\mathbf{r}, V)}{I(\mathbf{r}, -V)}|.
\end{align}
The Kondo lattice parameters used in Fig.~1 are taken from Refs.~\cite{Figgins2010, Figgins2011} with a slightly different Fermi wavelength of $8a$ instead of $10a$ to increase the number of oscillations in a fixed system size.

\newpage
\section{Supplementary Text}

\subsection{Rectification in Kondo lattice systems}

\begin{figure}[h!]
	\includegraphics[width=4.3in]{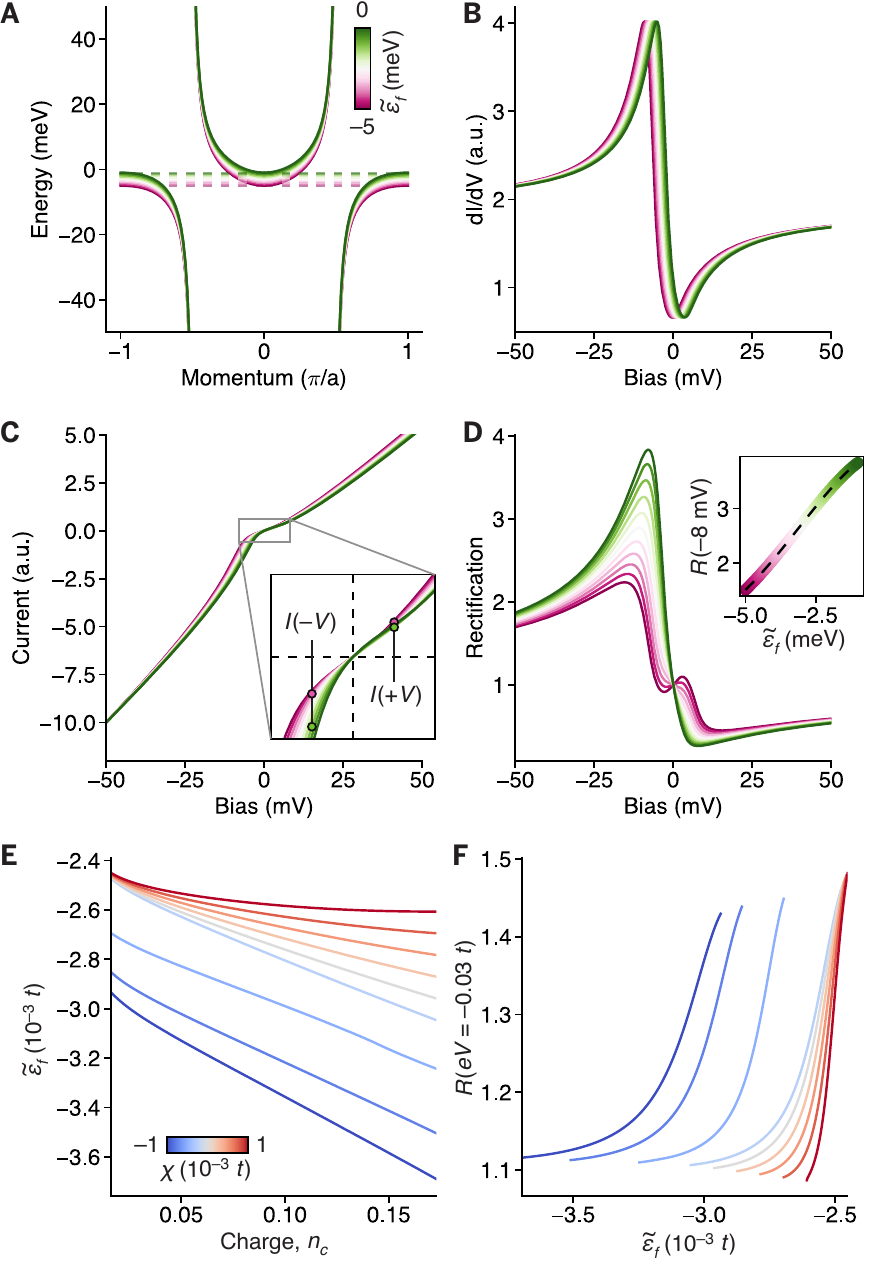}
		\caption{{\bf Rectification tracks variations in $\tilde{\varepsilon}_f$.}\label{fig:chi}
		({\bf A}) Tight-binding band structure of a Kondo lattice consisting of a parabolic conduction band with Fermi wavevector $k_F^c = 0.54\ \pi/a$ and minimum $-1.6$ eV (as measured by ARPES in SmB$_6$ \cite{Jiang2013}), and an $f$ band with dispersion $\chi=0$ meV, location $\tilde{\varepsilon}_f \in [-5, 0]$ meV and hybridization strength $\nu = 50$ meV. 
        ({\bf B}) Calculated differential tunneling conductance d$I$/d$V$ for different $\tilde{\varepsilon}_f$ keeping all other parameters fixed. For these calculations, we used STM tunneling amplitudes of $t_f /t_c = 0.03$ and an $f$-electron self energy of 3 meV. 
        ({\bf C}) Calculated current-voltage curve by integrating d$I$/d$V$ in panel (B). As the peak in d$I$/d$V$ shifts to lower biases (pink curves), the current at $+V$ becomes more similar to the current at $-V$. 
        ({\bf D}) Consequently, the peak in $R(V) = |I(+V)/I(-V)|$ at $V\approx -8$ mV decreases as $\tilde{\varepsilon}_f$ is reduced.
        ({\bf E}) In a self consistent solution to Eq.~\ref{eq:Hfull}, the local doping $n_c$ (set by adjusting $\mu$) is linearly related to the Kondo resonance position $\tilde{\varepsilon}_f$.
        ({\bf F}) The rectification tracks the changes in $\tilde{\varepsilon}_f$ with highest sensitivity (steepest slope) in the Kondo metal regime ($\chi > 0$).
	}
\end{figure}

\ptitle{} The dominant contribution to the local rectification in Kondo lattice systems arises from the large energy asymmetric d$I$/d$V$ peak caused by the Kondo resonance $\tilde{\varepsilon}_f$. As $\tilde{\varepsilon}_f$ moves in response to local charge, so does the energy position of the d$I$/d$V$ peak, which dramatically alters the odd component of d$I$/d$V$. The odd component of d$I$/d$V$ leads to an even component in $I(V)$, or an ``asymmetry'' since $I(V)$ is usually odd  (see Fig.~\ref{fig:chi}, A to D, for an illustration of this process):
    \begin{align}
        \frac{I(V)+I(-V)}{2} &= 
        \frac{1}{2} \int_0^V dV’ \frac{dI(V’)}{dV’} + \frac{1}{2} \int_0^{-V} dV’ \frac{dI(V’)}{dV’} \nonumber
        \\
        &= \frac{1}{2} \int_0^V dV’ \qty[\frac{dI(V’)}{dV’} - \frac{dI(-V’)}{dV’}]
    \end{align}
The rectification $R(V)$  can be written in terms of the $I(V)$ asymmetry as
    \begin{align}
        R(V) &= \frac{I(V)}{I(-V)} 
        = \frac{\int_0^V dV’ \frac{dI(V’)}{dV’}}{\int_0^V dV’ \frac{dI( -V’)}{dV’}}
        = \frac{1 + \alpha(V)}{1 - \alpha(V)}
    \end{align}
    where
    \begin{align}
       \alpha(V) &= \frac{\int_0^V dV’ \qty[\frac{dI(V’)}{dV’} - \frac{dI(-V’)}{dV’}]}{\int_0^V dV’\qty[\frac{dI(V’)}{dV’} + \frac{dI( -V’)}{dV’}]}
        = \frac{I(V) + I(-V)}{I(V) - I(-V)}
    \end{align}
Here $\alpha(V)$ is the ratio of the even and odd components of $I(V)$. As $\tilde{\varepsilon}_f$ moves towards the Fermi level, the odd component of d$I$/d$V$ grows quickly leading to an increase in $\alpha(V)$ and a large peak in $R(V)$. For pedagogical purposes, we isolated this mechanism in Fig.~\ref{fig:chi}, A to D, by artificially changing $\tilde{\varepsilon}_f$ and calculating the resulting d$I$/d$V$ with all other tight-binding parameters fixed. In a self-consistent solution to the model in Eq.~\ref{eq:Hfull}, we induce a change in $\tilde{\varepsilon}_f$ by adding additional charge $\Delta n_c$ to the system by adjusting $\mu$. In this case, we find that $\tilde{\varepsilon}_f$ depends linearly on $n_c$, and $R$ is a monotonic function of $\tilde{\varepsilon}_f$ that depends on the $f$-band dispersion $\chi(\vec{r}, \vec{r}')$, as shown in Fig.~\ref{fig:chi}, E and F.

\begin{figure}[h!]
	\includegraphics[width=5in]{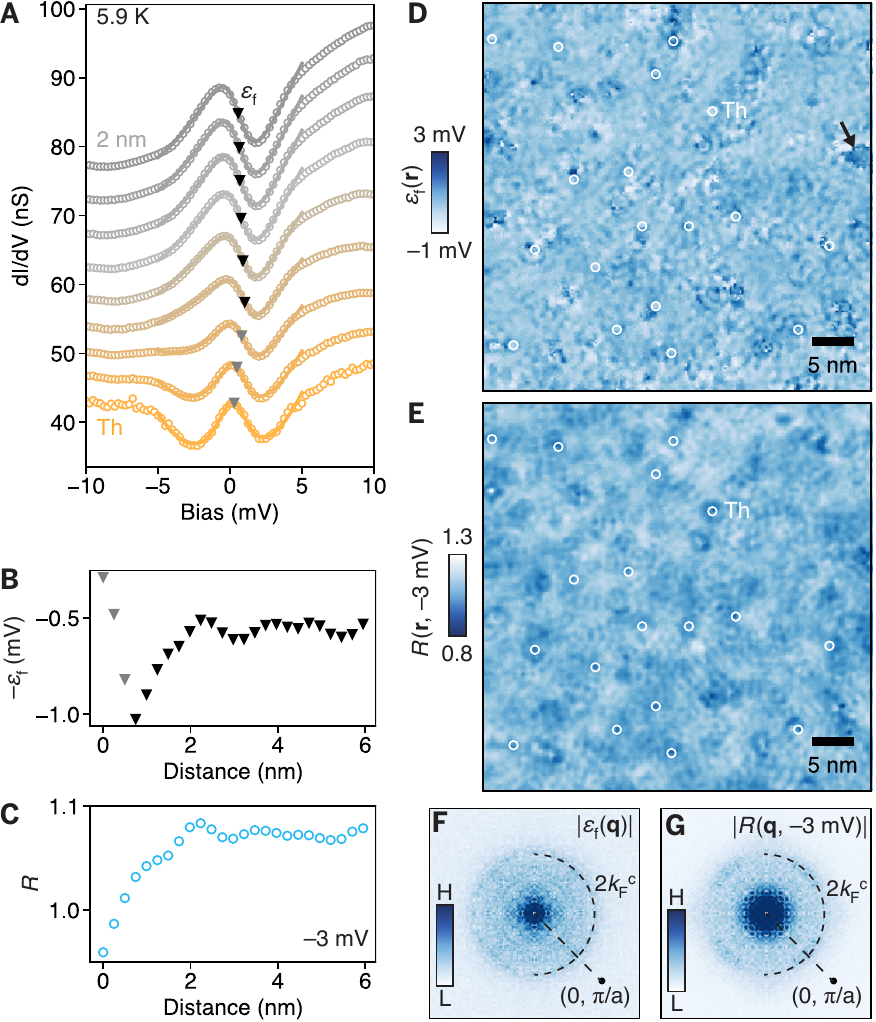}
	\caption{\label{fig:s1}{\bf $R(\mathbf{r}, V)$ tracks the position of the Kondo resonance in URu$_2$Si$_2$.}
    ({\bf A}) Measured d$I(\mathbf{r},V)$/d$V$ on the U termination of URu$_2$Si$_2$ (open circles, each curve offset for clarity) is well described by a Fano model plus a parabolic background (solid lines). As the tip approaches a Th dopant, the energy position of the Kondo resonance ($\varepsilon_f$, black triangles) shifts to maintain a constant $f$-electron occupancy. The spectra are averaged over the 18 well-isolated thorium dopants marked in (D). 
	({\bf B}) The Kondo resonance position extracted from the Fano model displays clear oscillations.
	({\bf C}) The simultaneously measured $R(\mathbf{r}> 1\ \mathrm{nm}, V=-3\ \mathrm{mV})$ accurately tracks the variations in $\varepsilon_f(\mathbf{r}> 1\ \mathrm{nm})$  [correlation coefficient: $R^2$ = –0.95, excluding gray points in (B)].
	({\bf D}) A map of $\varepsilon_f(\mathbf{r})$ extracted from fits to Eq.~\ref{eqn:s1} at each site contains clear ripples centered around Th sites, even though the fits fail in several small patches (e.g.~black arrow).
	({\bf E}) The measured $R(\mathbf{r}, V)$ (as shown in Fig.~2D) is highly sensitive to the Kondo resonance position.
	({\bf F-G}) Both measurements reveal oscillations at the parent Fermi wavevector $2k_\mathrm{F}^\mathrm{c} \approx 0.3\ (2\pi/a)$, indicating the presence of unhybridized electrons around Kondo holes. 
	}
\end{figure}

\ptitle{}In simple Kondo-lattice systems, d$I$/d$V$ often empirically follows a Fano function, a model typically used to describe the interference between a discrete state (here, the Kondo resonance) and a continuum (the conduction band), given by \cite{Schmidt2010, Giannakis2019,Figgins2019}
\begin{align}\label{eqn:s1}
    \frac{\mathrm{d}I(\vec{r}, V)}{\mathrm{d}V} \propto \frac{(E^\prime + q)^2}{E^\prime + 1} 
    \quad \mathrm{where} \quad
    E^\prime = \frac{eV - \varepsilon_f}{\Gamma/2}.
\end{align}
Here $\Gamma$ is the energy width of the Kondo resonance, $\varepsilon_f$ is its energy position, and $q$ is its asymmetry \cite{Figgins2019}. Where valid, Eq.~\ref{eqn:s1} allows $\varepsilon_f(\mathbf{r})$ to be extracted directly from d$I(\mathbf{r},V)$/d$V$, providing an independent check of the metallic $R(\mathbf{r},V)$ oscillations discovered in the main text. In  URu$_2$Si$_2$, d$I(\mathbf{r},V)$/d$V$ is adequately described by Eq.~\ref{eqn:s1} plus a parabolic background for low biases $|V|< 5$ mV (see Fig.~\ref{fig:s1}A). Near a Th dopant, the extracted $\varepsilon_f(\mathbf{r})$ trace displays clear oscillations that match those in $R(\mathbf{r},V)$ for $\mathbf{r} > 1$ nm, with a correlation coefficient of $R^2=-0.95$ (Fig.~\ref{fig:s1}, B and C). These two quantities differ at the defect site (where $\mathbf{r} < 1$ nm), likely because the validity of Eq.~\ref{eqn:s1} breaks down around Kondo holes. Nevertheless, the map of $\varepsilon_f(\mathbf{r})$ contains prominent oscillations at the parent-state Fermi wavevector of $2k_\mathrm{F}^\mathrm{c} \approx 0.3\ (2\pi/a)$ (see Fig.~\ref{fig:s1}, D and F), corroborating the existence of charge oscillations in $R(\mathbf{r},V)$ (Fig.~\ref{fig:s1}, E and G).

\subsection{Origin of $R(\mathbf{r}, V)$ oscillations}

\begin{figure}[h!]
	\includegraphics[width=5in]{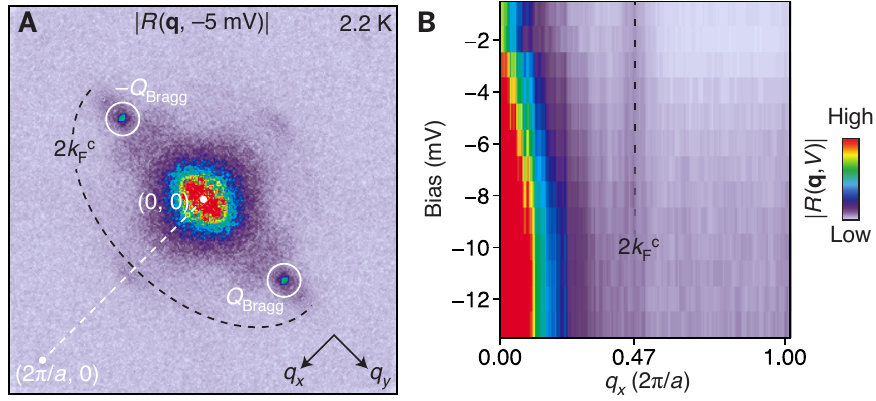}
	\caption{\label{fig:s2}{\bf Non-dispersive wavevector in $R(\mathbf{q}, V)$.}
    ({\bf A}) Measured $R(\mathbf{q}, V=-5\ \mathrm{mV})$ on SmB$_6$ (as shown in Fig.~3G) displays a sharp elliptical feature that has maximum contrast along the $\hat{q}_x$ direction (white dashed line). Conversely, the surface reconstruction creates a sharp peak along the $\hat{q}_y$ direction at $Q_\mathrm{Bragg} = (0, \pi/a)$.
	({\bf B}) Linecut of $R(\mathbf{q}, V)$ along $\hat{q}_x$ contains a peak at $2k_F^c$. This peak is dominant and non-dispersive for biases within the hybridization gap $|V| < \Delta/e \approx 10$ mV, but it is gradually washed out at higher $|V|$.
	}
\end{figure}

\ptitle{}The $R(\mathbf{r}, V)$ oscillations in SmB$_6$ have a different origin to the simultaneously measured heavy Dirac surface states that appear in our d$I$/d$V$ measurements. First, the largest major-axis wavevector of the surface state (see Fig.~4B) is about 30\% smaller than the extrapolated major axis of the $R(\mathbf{q})$ ellipse (dashed line in Fig.~\ref{fig:s2}A). Second, the surface state contributes very little quasiparticle interference at the bias corresponding to the Dirac point $V_D = E_D/e \approx -5$ mV, whereas the $R(\mathbf{q})$ ellipse is clear and prominent at this bias (see Fig.~\ref{fig:s2}A). Third, the surface state disperses rapidly for biases within the hybridization gap $|V| < \Delta/e \approx 10$ mV (see Fig.~4B), whereas the $R(\mathbf{q})$ ellipse is non dispersive (Fig.~\ref{fig:s2}B). Fourth, the rectification $R(V)$ from a Dirac-like density of states $\rho(E) \propto v_F |E - E_D|$ depends only on the Dirac point energy $E_D$. It has the explicit form 
    \begin{align}\label{eq:rect}
        R(V) = \begin{cases} 
            \frac{V(V-2V_D)}{(V+V_D)^2+V_D^2} & V V_D < - V_D^2 \\
            \frac{V-2V_D}{V+2V_D} & |V|\leq |V_D| \\
            \frac{(V-V_D)^2+V_D^2}{V(V+2V_D)} & V V_D > V_D^2 
            \end{cases}
    \end{align}
where $V_D = E_D/e$ is the bias corresponding to the Dirac point. In SmB$_6$, the Dirac point occurs slightly below $E_F$ at $E_D \approx -5$ meV (see Fig.~4B and \cite{Pirie2020}). Therefore, from Eq.~\ref{eq:rect}, the heavy Dirac surface states contribute a \textit{forward-biased} rectification curve with a peak at positive $V$, whereas our measured $R(V)$ curves reveal a \textit{reverse-biased} junction with a peak at negative $V$ (e.g.~Fig.~3C).

\begin{figure}[h!]
	\includegraphics[width=5in]{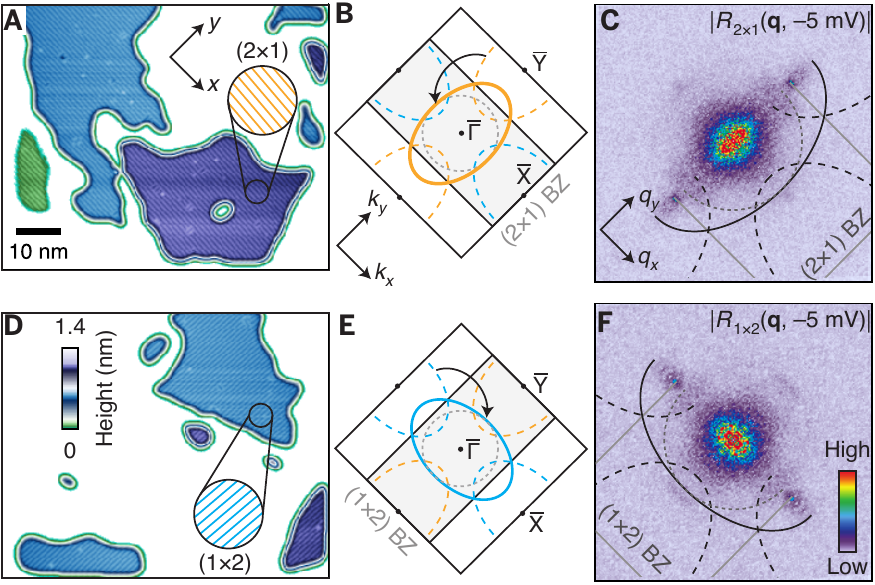}
		\caption{{\bf Possible wavevectors of the bulk $5d$ band on a half-Sm-terminated surface.}\label{fig:2x1}
        ({\bf A}) A $(2 \times 1)$ surface reconstruction doubles the size of the unit cell along the $\hat{y}$ direction, creating domains of uniform stripes as shown in this topography. We calculated the edges of the $(2 \times 1)$ domains from the inverse Fourier transform of the $(2 \times 1)$ Bragg peak. 
        ({\bf B}) Consequently, the Brillouin zone size is halved along $k_y$, folding the $\bar{Y}$ pocket to the $\bar{\Gamma}$ point.
        ({\bf C}) Measured $R(\mathbf{q})$ for the $(2 \times 1)$ domains shown in panel (B), overlaid with all possible $5d$ wavevectors. The signal we image most closely matches the $5d$ contour originating at the $\bar{Y}$ point and folded to $\bar{\Gamma}$ by the surface reconstruction (solid black line). 
        ({\bf D-F}) Same as (A-C) but for regions of the sample with a $(1 \times 2)$ surface reconstruction, which doubles the unit cell in the $\hat{x}$ direction. This reconstruction folds the $\bar{X}$ contour to the $\bar{\Gamma}$ point, such that the dominant $R(\mathbf{q})$ ellipse is oriented along $q_x$, perpendicular to that in panel (C).
	}
\end{figure}

\ptitle{}Although it cannot be explained by the surface state, the non-dispersive $R(\mathbf{q})$ ellipse matches expectations for the metallic $5d$ band, after accounting for its projection onto a $(2 \times 1)$ reconstructed surface. The surface projection of a bulk band typically traces the ($k_x, k_y$) contours of the corresponding bulk band for all $k_z$ points with zero group velocity in the $\hat{z}$ direction, $v_{\hat{z}} = \hbar^{-1}\grad_{\hat{k}_z}E(\mathbf{k})=0$. In SmB$_6$, this projection results in an ellipse at the $\bar{X}$ and $\bar{Y}$ points with $k_z=0$ and a circle around $\bar{\Gamma}$ with $k_z = \pi$. The $(2 \times 1)$ surface reconstruction reduces the size of the Brillouin zone in the $\hat{y}$ direction. This potential folds the $\bar{Y}$-point ellipse to the $\bar{\Gamma}$ point of the $(2 \times 1)$ surface Brillouin zone (see Fig.~\ref{fig:2x1}, B and E). The opposite effect occurs on $(1 \times 2)$ domains, which fold the $\bar{X}$-point ellipse to $\bar{\Gamma}$. We separated the field of view in Fig.~4A into regions of uniform $(2 \times 1)$ or $(1 \times 2)$ termination by inverse Fourier transforming the corresponding Bragg peak (Fig.~\ref{fig:2x1}, A and D). The $R(\mathbf{q})$ signal within each domain closely matches the folded $\bar{Y}$ or $\bar{X}$ ellipse (Fig.~\ref{fig:2x1}, C and F). In Fig.~3G, we averaged the $\bar{X}$ and the $\bar{Y}$ pockets by rotating the Fourier transform $R_{2\times 1}(\mathbf{q})$ by 90$^\circ$ before averaging with the Fourier transform $R_{1\times 2}(\mathbf{q})$.

\subsection{Metallic puddles around Gd dopants}

\begin{figure}[h!]
	\includegraphics[width=5in]{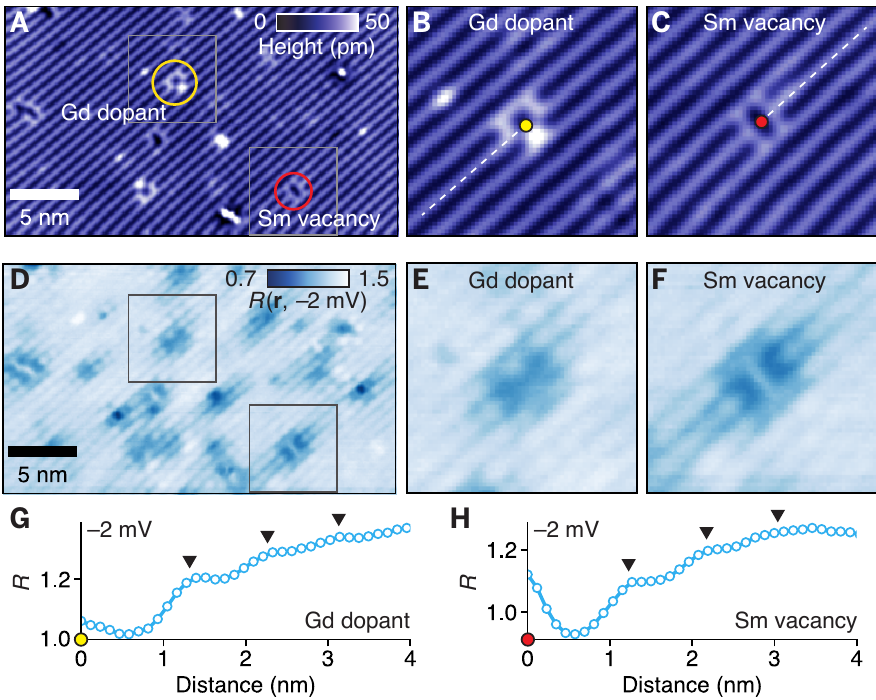}
	\caption{\label{fig:s3}{\bf Metallic puddles around Gd dopants in SmB$_6$.}
    ({\bf A-C}) Topography of the $(2\times 1)$-reconstructed Sm termination of an SmB$_6$ sample lightly doped with Gd dopants (e.g.~yellow circle), in addition to native defects like Sm vacancies (red circle).
	({\bf D-F}) Measured $R(\mathbf{r}, V)$ displays similar oscillations around Gd dopants as other Sm-site defects, such as Sm vacancies.
	({\bf G-H}) Linecut of $R(\mathbf{r}, V=-2\ \mathrm{mV})$ around ({\bf G}) a Gd dopant along the dashed line in (B) and ({\bf H}) a Sm vacancy along the dashed line in (C). 
	}
\end{figure}

\ptitle{} To verify the discovery of metallic puddles in SmB$_6$, we measured $R(\mathbf{r},V)$ around a third common Sm-site defect: Gd dopants. Gd dopants are suspected to form spinful Kondo holes in SmB$_6$ because they are known to dope magnetically \cite{Fuhrman2018}. They also contribute strongly to the residual linear specific heat \cite{Fuhrman2018} and appear to generate local metallic puddles from electron-spin resonance measurements \cite{Souza2020}. Our $R(\mathbf{r},V)$ data confirms the existence of these metallic puddles, as we see the same unhybridized $5d$ wavevector around Gd dopants as we do around Sm vacancies (compare both dopants in Fig.~\ref{fig:s3}, E and F). The similarity between the $R(\mathbf{r})$ maps around each dopant type, despite their different topographies, implies that $R(\mathbf{r})$ couples primarily to the parent Fermi surface and not to a local property of the specific dopant (such as an impurity bound state). This similarity also indicates that the defect form factor does not significantly impact the $R(\mathbf{r})$ maps.

\subsection{Quantum oscillations from metallic puddles}

\begin{figure}[h!]
	\includegraphics[width=5in]{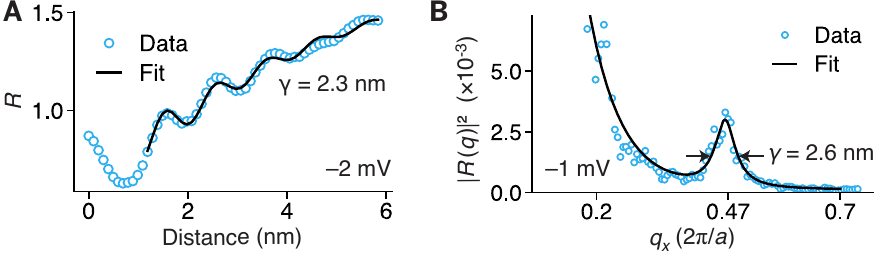}
		\caption{\label{fig:size}
    	{\bf Size of the metallic puddles in SmB$_6$.}
        ({\bf A}) The oscillations in $R(\mathbf{r})$ around an Fe dopant have an estimated decay length of $\gamma=2.3$ nm, calculated by fitting the $R(\mathbf{r})$ linecut (blue points) to a decaying oscillation $\propto e^{–x/\gamma} \cos(kx)$ plus a quadratic background (black line).
        ({\bf B}) In a larger field of view containing 15 well-isolated Kondo holes, the average decay length is $\gamma=2.6$ nm, calculated by fitting $|R(q_x)|^2$ data to a Lorentzian peak plus an exponential background. 
	}
\end{figure}

\ptitle{} Our discovery of light $d$ electrons around point defects suggests an alternative origin for the measured de Haas-van Alphen (dHvA) effect in SmB$_6$. These metallic puddles are large enough to host Landau levels at the magnetic field of 35 T where high-frequency (large-$k_F$) dHvA oscillations are first observed \cite{Hartstein2018}. Specifically, the real-space area enclosed by a Landau orbit is $S = \pi l_B^2$, where $l_B = \sqrt{\hbar/eB} \approx 4.3$ nm is the magnetic length at 35 T, whereas the puddle radius is typically a few times the $R(\mathbf{r})$ decay length of $\gamma = 2.6$ nm (see Fig.~\ref{fig:size}).

\ptitle{} Our discovery also reveals the theoretical challenge to model the emergence of dHvA oscillations from disordered metallic puddles. For example, a confining potential is known to qualitatively modify the dHvA response \cite{Curnoe1998}, changing both the shape and amplitude of the oscillations \cite{Herzog2016}, and typically violating Lifshitz-Kosevich theory \cite{Champel2001}. In SmB$_6$, the dHvA amplitude was already shown to deviate from Lifshitz-Kosevich behavior at low temperatures \cite{Hartstein2018}. Additionally, the onset of the large-$k_F$ oscillations at 35 T violates the traditional low-scattering condition $\omega_c \tau  = l/r_c \gg 1$, as the estimated mean free path of $l \approx $ 10-50 nm is smaller than the cyclotron radius $r_c \approx 109$ nm \cite{Hartstein2018}. An interesting and open question is whether the Landau orbits are really confined to a single puddle, or whether the overlap between puddles is sufficient to generate orbits that involve multiple puddles. Bulk measurements of floating-zone-grown SmB$_6$ suggest a vacancy concentration of about $d=1\%$, translating to an average nearest-neighbor separation of $a d^{-\frac{1}{3}} \approx 2$ nm, where $a=0.413$ nm is the SmB$_6$ lattice constant. This separation is comparable to the measured $R(\mathbf{r})$ decay length of $\gamma = 2.6$ nm, suggesting some samples may contain regions of extended but non-percolating metallic clusters. Predicting the dHvA response of such a spatially heterogeneous and strongly interacting system is a pertinent and pressing theoretical task.

\newpage

\newpage

\newpage

\newpage

\newpage

\newpage

\newpage

\newpage
\begin{table}[h!]
    \caption{Comparison of nominal and measured doping levels for the three samples analyzed in the main text.
        }\label{tab:doping}
    \begin{ruledtabular}
    \begin{tabular}[c]{lccc}
        Sample  & Nominal  & Measured  &  Sm vacancies
        \\ \hline
        Th-URu$_2$Si$_2$    &  1\%  & 0.5\%  &  -
        \\
        Gd-SmB$_6$          & 0.3\% & 0.15\% &  0.33\%
        \\
        Fe-SmB$_6$          & 0.5\% & 0.05\% &  0.09\%
        \\
    \end{tabular}
    \end{ruledtabular}
\end{table}

\newpage
\bibliography{refs}